\documentclass[10pt,conference]{IEEEtran}

\usepackage{research5}

\newtheorem{dfn}{Definition}
\newtheorem{thm}{Theorem}
\newtheorem{cor}{Corollary}
\newtheorem{lm}{Lemma}
\newtheorem{assumption}{Assumption}

\begin{document}

\title{Sending a Bivariate Gaussian Source over a
  Gaussian MAC with Feedback}

\author{\authorblockN{Amos Lapidoth ~ ~ ~ Stephan Tinguely}
\authorblockA{Signal and Information Processing Laboratory\\
ETH Zurich, Switzerland\\
\texttt{ \{lapidoth, tinguely\}@isi.ee.ethz.ch}}}

\maketitle

\begin{abstract}
\renewcommand{\thefootnote}{}
We consider the problem of transmitting a bivariate Gaussian source
over a two-user additive Gaussian multiple-access channel with
feedback. Each of the transmitters observes one of the source
components and tries to describe it to the common receiver. We are
interested in the minimal mean squared error at which the receiver can
reconstruct each of the source components.

In the ``symmetric case'' we show that, below a certain signal-to-noise
ratio threshold which is determined by the source correlation,
feedback is useless and the minimal distortion is achieved by uncoded
transmission. For the general case we give necessary conditions for
the achievability of a distortion pair.\footnote{The work of Stephan Tinguely
  has been partially supported by the Swiss National Science Foundation
  under Grant 200021-111863/1.} \setcounter{footnote}{0}
\end{abstract}

\section{Introduction}
We consider the problem of transmitting a memoryless bivariate
Gaussian source over a two-user additive white Gaussian
multiple-access channel with perfect causal feedback from the channel
output to both transmitters. Each of the transmitters observes,
besides the previous channel outputs, one of the source components
which it tries to describe to the receiver subject to an average power
constraint on its transmitted signal. Based on the channel output, the
receiver estimates the two source components. The quality of the
estimate is measured in squared-error distortion on each individual
component. We seek the achievable distortion pairs.

We show that in the ``symmetric case'' --- where the transmitters are
subjected to the same average power constraint and the ratio of the
distortions to be achieved is equal to the ratio of the corresponding
source variances --- there is a threshold signal-to-noise ratio (SNR),
determined by the correlation between the source components, below
which feedback is useless and the minimal distortion is achieved by
uncoded transmission. This result strengthens a previous result of
Lapidoth and Tinguely \cite{lapidoth-tinguely06} for the same problem
but without feedback. For the general case we give necessary conditions
for the achievability of a distortion pair.

Related results by Oohama \cite{oohama97} and Wagner et
al.~\cite{wagner-tavildar-vishwanath05} only treated the source coding
aspect of this problem by solving the Slepian-Wolf lossy version for
the bivariate Gaussian source and by Ozarow \cite{ozarow85} who only
treated the channel coding aspect by computing the capacity region of
the Gaussian multiple-access channel with feedback. We shall,
however, not rely on these source coding and channel coding results
since the separation theorem does not apply to our problem. That
feedback is useless in the symmetric case below some threshold SNR is
all the more surprising in view of the recent work of Lapidoth and
Wigger \cite{lapidoth-wigger06} who showed that feedback, even if
noisy, always increases the capacity region of the Gaussian
multiple-access channel.

\section{Problem Statement}

We consider a discrete-time two-user additive white Gaussian
multiple-access channel with perfect and causal feedback from the
channel output to both transmitters. The two transmitters of the
multiple-access channel each observe one component of a memoryless
bivariate Gaussian source and try to communicate it to the receiver.

The time-$k$ output of the Gaussian multiple-access channel is given by
\begin{equation}
Y_k = x_{1,k} + x_{2,k} + Z_k,
\end{equation}
where $x_{1,k} \in \Reals$ and $x_{2,k} \in \Reals$ are the symbols
sent by the two transmitters, and $Z_k$ is the time-$k$ additive
noise term. The terms $\{Z_{k}\}$ are independent identically
distributed (IID) zero-mean variance-$N$ Gaussian random variables
that are independent of the source sequence.

The source symbols produced at time $k$ are $(S_{1,k}, S_{2,k})$ where
the $\{ (S_{1,k}, S_{2,k}) \}$ are IID zero-mean Gaussians of
covariance
\begin{equation}\label{eq:source-law}
\cov{S} = \left( \begin{array}{c c}
\sigma_1^2 & \rho \sigma_1 \sigma_2\\
\rho \sigma_1 \sigma_2 & \sigma_2^2
\end{array} \right),
\end{equation}
with $\rho \in [-1,1]$, and $0 < \sigma_i^2 < \infty$, $i = 1,2$. The
sequence of the first source component $\{S_{1,k}\}$ is observed by
Transmitter~1 and the sequence of the second source component
$\{S_{2,k}\}$ is observed by Transmitter~2. Based on their source
sequence and the feedback observed up to time $k$, the transmitters
produce their respective time-$k$ channel inputs
\begin{displaymath}
x_{i,k} = f_{i,k}^{(n)} \left( {\bf S}_i, Y^{k-1} \right) \qquad i=1,2,
\end{displaymath}
where we have used the shorthand notation ${\bf S}_i = (S_{i,1}, \ldots
,S_{i,n})$ and $Y^{k-1} = (Y_1, \ldots ,Y_{k-1})$, and where
\begin{equation}\label{eq:encoder}
  f_{i,k}^{(n)} : \Reals^{n} \times \Reals^{k-1} \rightarrow
  \Reals, \qquad i=1,2, \quad k=1,\ldots ,n.
\end{equation}
The transmitted sequences of the two encoders are average-power
limited to $P_{1}$ and $P_{2}$ respectively, i.e.
\begin{equation}\label{eq:power}
  \frac{1}{n} \sum_{k=1}^n\E{ \left(
      f_{i,k}^{(n)}\bigl({\bf S}_i,Y^{k-1} \bigr) \right)^{2}} \leq
  P_{i}, \quad i=1,2.
\end{equation}
The decoder estimates the two source sequences based on the channel
output ${\bf Y} = (Y_1, \ldots Y_n)$. These estimates are denoted by
$\hat{\bf S}_1 = \phi_1^{(n)}({\bf Y})$ and $\hat{\bf S}_2 =
\phi_2^{(n)}({\bf Y})$ respectively, where
\begin{equation}\label{eq:reconstructor}
  \phi_{i}^{(n)} : \Reals^{n} \rightarrow \Reals^{n}, \quad i=1,2.
\end{equation}
We are interested in the minimal expected squared-error distortions at
which the receiver can reconstruct each of the source sequences.

\begin{dfn}
Given $\sigma_{1}, \sigma_{2} > 0$, $\rho \in [-1,1]$, $P_{1}, P_{2} >
0$, and $N > 0$ we say that the tuple $(D_{1}, D_{2},
\sigma^{2}_{1}, \sigma_{2}^{2}, \rho, P_{1}, P_{2}, N )$ is
\emph{achievable} if there exists a sequence of encoding functions
$(f_{1,k}^{(n)}, f_{2,k}^{(n)})$ as in (\ref{eq:encoder}) and a
sequence of reconstruction pairs $(\phi_{1}^{(n)}, \phi_{2}^{(n)})$ as
in (\ref{eq:reconstructor}) satisfying the average power constraints
(\ref{eq:power}) and resulting in average distortions that fulfill
\begin{displaymath}
  \varlimsup_{n \rightarrow \infty} \frac{1}{n} \sum_{k=1}^n \E{
    \left( S_{i,k} - \hat{S}_{i,k} \right)^2} \leq D_{i}, \quad i=1,2,
\end{displaymath}
whenever
\begin{displaymath}
Y_k = f_{1,k}^{(n)}({\bf S}_1,Y^{k-1}) + f_{2,k}^{(n)}({\bf S}_2,Y^{k-1}) +
Z_k,
\end{displaymath}
for $k = 1,\ldots n,$ and $\{(S_{1,k},S_{2,k})\}$ are IID zero-mean
bi-variate Gaussian vectors of covariance matrix $\cov{S}$ as in
(\ref{eq:source-law}) and $\{Z_{k}\}$ are IID zero-mean variance-$N$
random variables that are independent of $\{(S_{1,k},S_{2,k})\}$.
\end{dfn}

The problem we address here is, for given $\sigma_1^{2}$,
$\sigma_2^{2}$, $\rho$, $N$, $P_{1}$, $P_{2}$, to find the set of
pairs $(D_{1}, D_{2})$ such that $(D_{1}, D_{2}, \sigma_1^{2},
\sigma_2^{2}, \rho, P_{1}, P_{2}, N)$ is achievable.

{\bf Remark:} As in \cite[Section III]{lapidoth-tinguely06} it can be
shown that there is no loss in generality in assuming that the two
source components are of equal variance and that the correlation
coefficient is non-negative. Hence, for the remainder we shall assume
\begin{displaymath}
\rho \in [0,1] \qquad \text{and} \qquad \sigma_1^2 =
\sigma_2^2 = \sigma^2.
\end{displaymath}
Furthermore, the convexity argument of \cite[Section
III]{lapidoth-tinguely06} applies also to the case with
feedback so that for any given $\sigma^2$, $\rho$, and $N$, the set of
all $(D_1,D_2,P_1,P_2)$ such that
$(D_1,D_2,\sigma^2,\sigma^2,\rho,P_1,P_2,N)$ is achievable is a convex set.

Of special interest is the ``symmetric case'' of this problem where
both transmitters are subject to equal power constraints, and where we
seek to achieve the same distortion on each source component. That is,
for some given $N$ and $P_{1} = P_{2} = P$ we are interested in 
\begin{multline*}
 D^{*}(\sigma^{2}, \rho, P, N) \triangleq \inf \{ \max \{ D_1,D_2 \}:\\
(D_1, D_2, \sigma^{2}, \sigma^{2}, \rho, P, P, N) \text{ is achievable}\}.
\end{multline*}

\section{Main Results}

We now present necessary conditions for the achievability of
$(D_1,D_2,\sigma^2,\sigma^2,\rho,P_1,P_2, N)$ and show that in the
symmetric case if $P/N \leq \rho/(1-\rho^2)$ then the minimal distortion
$D^{\ast}(\sigma^2,\rho,P,N)$ is achieved by uncoded transmission and
feedback is useless. The corresponding proofs will be discussed in
Section \ref{sec:proofs}.

Denote by $R_{S_1,S_2}(D_1,D_2)$ the rate-distortion function for the
pair $(S_1,S_2)$ when this pair is observed by one common encoder. For
$(S_1,S_2)$ jointly Gaussian as in (\ref{eq:source-law}) and with
$\sigma_1^2 = \sigma_2^2 = \sigma^2$, we have
\begin{multline}\label{eq:Rd-S1-S2}
R_{S_1,S_2}(D_1,D_2)\\[3mm]
=\left\{ \begin{array}{l}
\frac{1}{2} \log_2 \left( \frac{\sigma^4(1-\rho^2)}{D_1D_2} \right)
\qquad \qquad \quad \text{if }(D_1,D_2)
\in \mathscr{D}_a\\[3mm]
\frac{1}{2} \log_2 \left(
  \frac{\sigma^4(1-\rho^2)}{D_1D_2 - \left( \rho \sigma^2 -
      \sqrt{(\sigma^2 - D_1)(\sigma^2 - D_2)} \right)^2 }\right) \\[5mm]
\qquad \qquad \qquad \qquad \qquad \qquad \; \: \text{if }(D_1,D_2) \in
\mathscr{D}_b\\[1mm]
\frac{1}{2} \log_2 \left( \frac{\sigma^2}{D_1}
\right) \qquad \qquad \qquad \quad \, \text{if }(D_1,D_2) \in \mathscr{D}_c,
\end{array}
\right.
\end{multline}
where the regions $\mathscr{D}_a$, $\mathscr{D}_b$ and $\mathscr{D}_c$
are given by 
\begin{align*}
\mathscr{D}_a &= \left\{ D_1 \leq \sigma^2 (1-\rho^2), D_2 \leq
  (\sigma^2(1-\rho^2)-D_1) \frac{\sigma^2}{\sigma^2-D_1} \right\}\\[3mm]
\mathscr{D}_b &= \bigg\{ 0 \leq D_1 \leq \sigma^2,\\
&(\sigma^2(1-\rho^2) - D_1)\frac{\sigma^2}{\sigma^2-D_1} \leq D_2 \leq
\sigma^2(1-\rho^2) + \rho^2 D_1 \bigg\}\\[3mm]
\mathscr{D}_c &= \left\{ 0 \leq D_1 \leq \sigma^2, D_2 >
  \sigma^2(1-\rho^2) + \rho^2D_1 \right\}.
\end{align*}
The expression for $R_{S_1,S_2}(D_1,D_2)$ has been derived in
\cite{xiao-luo05} and \cite{lapidoth-tinguely06} by different
approaches.

Further, denote by $R_{S_1|S_2}(D_1)$ the rate-distortion function for
$S_1$, when $S_2$ is known to both, the encoder and the decoder, and
analogously by $R_{S_2|S_1}(D_2)$ the rate-distortion function for
$S_2$, when $S_1$ is known to both, the encoder and the decoder. For
$(S_1,S_2)$ jointly Gaussian as in (\ref{eq:source-law}) and with
$\sigma_1^2 = \sigma_2^2 = \sigma^2$, we have
\begin{align}
R_{S_1|S_2}(D_1) &= \frac{1}{2} \log_2 \left(
  \frac{\sigma^2(1-\rho^2)}{D_1}\right) \label{eq:Rd-S1|S2} \\[3mm]
R_{S_2|S_1}(D_2) &= \frac{1}{2} \log_2 \left(
  \frac{\sigma^2(1-\rho^2)}{D_2}\right). \label{eq:Rd-S2|S1}
\end{align}

\begin{thm}\label{thm:general-bound}
A necessary condition for the achievability of
$(D_1,D_2,\sigma^2,\sigma^2,\rho,P_1,P_2, N)$ is that there exists a
$\tilde{\rho} \in [0,1]$ such that
\begin{align}
R_{S_1,S_2}(D_1,D_2) &\leq \frac{1}{2} \log_2 \left( 1 +
  \frac{P_1+P_2+2\tilde{\rho} \sqrt{P_1P_2}}{N}\right) \label{eq:thm-cond1}\\[2mm]
R_{S_1|S_2}(D_1) &\leq \frac{1}{2} \log_2 \left( 1 +
  \frac{P_1(1-\tilde{\rho}^2)}{N} \right) \label{eq:thm-cond2}\\[2mm]
R_{S_2|S_1}(D_2) &\leq \frac{1}{2} \log_2 \left( 1 +
  \frac{P_2(1-\tilde{\rho}^2)}{N} \right), \label{eq:thm-cond3}
\end{align}
where the explicit forms of the rate-distortion functions on the LHS,
are given in (\ref{eq:Rd-S1-S2}), (\ref{eq:Rd-S1|S2}), and
(\ref{eq:Rd-S2|S1}) respectively.  
\end{thm}

In the symmetric case, (\ref{eq:thm-cond1}) \& (\ref{eq:Rd-S1-S2})
yield
\begin{equation}\label{eq:sym-cond1}
D \geq \left\{ \begin{array}{l l}
\frac{1}{2} \left( \frac{N \sigma^2 (1+\rho)}{N+2P(1+\tilde{\rho})} +
  \sigma^2(1-\rho)\right) & \text{if }\frac{P}{N} \leq
\frac{\rho}{1-\rho^2}\\[3mm]
\sigma^2 \sqrt{ \frac{N (1-\rho^2)}{N+2P(1+\tilde{\rho})}} & \text{if
}\frac{P}{N} > \frac{\rho}{1-\rho^2}
\end{array} \right.
\end{equation}
and (\ref{eq:thm-cond2}) \& (\ref{eq:Rd-S1|S2}) (or
(\ref{eq:thm-cond3}) \& (\ref{eq:Rd-S2|S1})) yield
\begin{equation}\label{eq:sym-cond2}
D \geq \sigma^2 \frac{N (1-\rho^2)}{N+P(1-\tilde{\rho}^2)}.
\end{equation}
We denote the RHS of (\ref{eq:sym-cond1}) by $\xi (\sigma^2, \rho, P,
N, \tilde{\rho})$ and the RHS of (\ref{eq:sym-cond2}) by $\psi
(\sigma^2, \rho, P, N, \tilde{\rho})$. 

\begin{cor}\label{cor:symmetric-case}
In the symmetric case
\begin{multline*}
D^{\ast}(\sigma^2,\rho,P,N) \geq\\ \min_{0 \leq \tilde{\rho} \leq 1}
\max \left\{ \xi (\sigma^2, \rho, P, N, \tilde{\rho}), \psi (\sigma^2,
  \rho, P, N, \tilde{\rho}) \right\}.
\end{multline*}
\end{cor}
\vspace{4mm}
\textbf{Note:} For $P/N \leq \rho^2/(2(1-\rho)(1+2\rho))$ the minimum
in Corollary \ref{cor:symmetric-case} is achieved by $\tilde{\rho} =
1$, and for all larger $P/N$ the minimum is achieved by the
$\tilde{\rho}^{\ast}$ for which
\begin{displaymath}
\xi (\sigma^2, \rho, P, N, \tilde{\rho}^{\ast}) = \psi (\sigma^2,
\rho, P, N, \tilde{\rho}^{\ast}).
\end{displaymath}

We can now verify that for $P/N = \rho/(1-\rho^2)$ the lower bound on
$D^{\ast}(\sigma^2,\rho,P,N)$ from Corollary \ref{cor:symmetric-case}
is achieved by uncoded transmission.
For $P/N = \rho/(1-\rho^2)$ the minimizing
$\tilde{\rho}$ is $\tilde{\rho}^{\ast} = \rho$ leading to the bound
\begin{equation}\label{eq:one-point}
D^{\ast}(\sigma^2,\rho,P,N) \geq \sigma^2 (1-\rho).
\end{equation}

To see that this is achievable by uncoded transmission, note that
in the symmetric case, uncoded transmission of the form  $x_{i,k} =
\sqrt{P/\sigma^2} S_{i,k}$, $i=1,2$ results in the distortion
\begin{equation}\label{eq:uncoded}
D_u \triangleq \sigma^2 \frac{P(1-\rho^2) +N}{2P(1+\rho)
  +N},
\end{equation}
(see \cite[Corollary 2]{lapidoth-tinguely06}), which, when evaluated
at $P/N = \rho/(1-\rho^2)$ yields the RHS of (\ref{eq:one-point}). The
following theorem extends this result to all $P/N \leq \rho/(1-\rho^2)$.
\begin{thm}\label{thm:symmetric-case}
In the symmetric case if $P/N \leq \rho/(1-\rho^2)$ we have
\begin{equation}\label{eq:thm-symmetric}
D^{\ast}(\sigma^2,\rho,P,N) = \sigma^2 \frac{P(1-\rho^2) +
  N}{2P(1+\rho) + N},
\end{equation}
i.e.~the minimal distortion is achieved by uncoded transmission, and
the availability of feedback is useless.
\end{thm}

\section{Sketches of Proofs}\label{sec:proofs}

We shall discuss the proofs of both theorems but with more
particularity on the proof of Theorem \ref{thm:symmetric-case}. We do
so, because the basic techniques to the proof of Theorem
\ref{thm:general-bound} are the same as in \cite{ozarow85} and
\cite[page 15]{gastpar-thesis}.

\subsection{Proof of Theorem \ref{thm:general-bound}}
To prove Theorem \ref{thm:general-bound} we shall use the following
lemma
\begin{lm}\label{lemma:MAC-rate}
Let the sequences $\{ X_{1,k} \}$ and $\{ X_{2,k} \}$ satisfy
$\sum_{i=1}^n \E{X_{i,k}^2} \leq n P_i$, $i=1,2$. Let $Y_k = X_{1,k} +
X_{2,k} + Z_k$, where $\{Z_k\}$ are IID zero-mean variance-$N$
Gaussian, and where for every $k$, $Z_k$ is independent of $(X_{1,k},
X_{2,k})$. Let $\tilde{\rho} \in [0,1]$ be given by
\begin{equation}\label{eq:tilde-rho}
\tilde{\rho} = \frac{\left| \frac{1}{n} \sum_{k=1}^n
    \E{X_{1,k}X_{2,k}} \right|}{\sqrt{\left( \frac{1}{n} \sum_{k=1}^n
      \E{X_{1,k}^2} \right) \left( \frac{1}{n} \sum_{k=1}^n
      \E{X_{2,k}^2} \right)}}.
\end{equation}
Then
\begin{align}
\sum_{k=1}^n I(X_{1,k},X_{2,k} ; Y_k) &\leq \frac{n}{2} \log_2 \left( 1 +
  \frac{P_1 + P_2 + 2 \tilde{\rho} \sqrt{P_1P_2}}{N} \right), \label{eq:MAC-sum-rate}\\
\sum_{k=1}^n I(X_{1,k} ; Y_k|X_{2,k}) &\leq \frac{n}{2} \log_2 \left( 1
  + \frac{P_1(1-\tilde{\rho}^2)}{N} \right),\label{eq:MAC-rate1}\\
\sum_{k=1}^n I(X_{2,k} ; Y_k|X_{1,k}) &\leq \frac{n}{2} \log_2 \left( 1
  + \frac{P_2(1-\tilde{\rho}^2)}{N} \right).\label{eq:MAC-rate2}
\end{align}
\end{lm}
\vspace{4mm}

The proof of Lemma \ref{lemma:MAC-rate} follows from the proof of the
main result in \cite{ozarow85} and is omitted. Theorem
\ref{thm:general-bound} can now be proved by showing
\begin{align}
n R_{S_1,S_2}(D_1,D_2) &\leq I({\bf S}_1,{\bf S}_2;{\bf Y})
\label{eq:prf-sum-rate1} \\ 
I({\bf S}_1,{\bf S}_2;{\bf Y}) &\leq \sum_{k=1}^n
I(X_{1,k},X_{2,k};Y_k), \label{eq:prf-sum-rate2} \\[2mm]
n R_{S_1|S_2}(D_1) &\leq I({\bf S}_1;{\bf Y}|{\bf S}_2)
\label{eq:prf-side-info1-1} \\
I({\bf S}_1;{\bf Y}|{\bf S}_2) &\leq \sum_{k=1}^n I(X_{1,k};Y_k|X_{2,k}),
\label{eq:prf-side-info1-2} \\[2mm]
n R_{S_2|S_1}(D_2) &\leq I({\bf S}_2;{\bf Y}|{\bf S}_1)
\label{eq:prf-side-info2-1} \\
I({\bf S}_2;{\bf Y}|{\bf S}_1) &\leq \sum_{k=1}^n I(X_{2,k};Y_k|X_{1,k}),
\label{eq:prf-side-info2-2}
\end{align}
and by then jointly bounding the expressions on the RHS of
(\ref{eq:prf-sum-rate2}), (\ref{eq:prf-side-info1-2}), and
(\ref{eq:prf-side-info2-2}) by means of Lemma
\ref{lemma:MAC-rate}. The proofs of (\ref{eq:prf-sum-rate1}) --
(\ref{eq:prf-side-info2-2}) follow along the same lines as the proof of
the univariate analog of which the derivations can be found in
\cite[page 15]{gastpar-thesis} (also coarsely stated in \cite[equation
(8)]{gastpar-rimoldi-vetterli03}). The main ingredients in those
derivations are the convexity of the rate-distortion functions and the
data-processing inequality. $\Box$

\subsection{Proof of Theorem \ref{thm:symmetric-case}}
To prove the theorem we need to show that $D^{\ast} \geq D_{u}$
whenever $P/N \leq \rho/(1-\rho^{2})$, where $D^{\ast}$ is short for 
$D^{\ast}(\sigma^2,\rho,P,N)$. Since the optimal reconstruction is the
conditional expectation, it suffices that we show that a contradiction
arises from the assumption:
\begin{assumption}[Leading to a contradiction]
\label{as:false}
The encoding rules $\{f_{i,k}^{(n)}\}$ satisfy the average
power constraints \eqref{eq:power} for some $P_{1}=P_{2}=P$ satisfying
$P/N \leq \rho/(1- \rho^{2})$
and, when combined with the optimal conditional expectation
reconstructors, achieve $D^{\ast}$, where $D^{\ast} < D_{u}$
\end{assumption}
To show that this assumption leads to a contradiction, let $\{
X_{1,k}, X_{2,k} \}$ and $\{ Y_k \}$ be the resulting channel inputs
and channel outputs when $\{f_{i,k}^{(n)}\}$ are used to describe the
source. Let further $\hat{\bf S}_1 = \E{{\bf S}_1 | {\bf Y}}$ and
$\hat{\bf S}_2 = \E{{\bf S}_2 | {\bf Y}}$. 

We focus on the estimation that Transmitter~2 can make for the vector
${\bf W} \triangleq {\bf S}_1 - \rho {\bf S}_2$ using his knowledged
of ${\bf S}_{2}$ and (through the feedback link) ${\bf Y}$. This
vector is the part of $({\bf S}_1,{\bf S}_2)$ which is independent of
${\bf S}_2$ and hence initially completely unknown to Transmitter 2.
However, from the feedback link Transmitter 2 can retrieve information
about ${\bf W}$.  The contradiction we shall obtain will be on the
distortion on ${\bf W}$ that can be achieved at Transmitter 2. Under
Assumption~\ref{as:false}, we shall derive contradictory lower and
upper bounds on the achievable value for this distortion.

For any estimator $\varphi^{(n)} ({\bf S}_2,{\bf Y})$ we set
\begin{displaymath}
D_W(\varphi^{(n)}) \triangleq \frac{1}{n} \E{\| {\bf W} - \varphi^{(n)}({\bf
    S}_2,{\bf Y}) \|^2},
\end{displaymath}
where $\| {\bf v} \|^2 = \sum_{k=1}^n v_k^2$.

\subsubsection{``Lower Bound'' on $D_W(\varphi^{(n)})$}
In this subsection we show that
\begin{multline}\label{eq:lb-Dw}
\text{Assumption \ref{as:false} } \Rightarrow\\
D_W(\varphi^{(n)}) > \sigma^2 (1-\rho^2) \frac{N}{N + P(1-\rho^2)}
\; \; \; \forall \varphi^{(n)}.
\end{multline}
The main ingredient is the following lemma:
\begin{lm}\label{lemma:mut-info}
\begin{multline*}
\text{Assumption \ref{as:false} } \Rightarrow\\
I({\bf S}_1;{\bf Y} | {\bf S}_2) < \frac{n}{2} \log_2 \left( 1 +
  \frac{P(1-\rho^2)}{N} \right). 
\end{multline*}
\end{lm}
\vspace{3mm}
The proof of Lemma \ref{lemma:mut-info} will be discussed in Section
\ref{subsec:proofs-lemmas}. Inequality (\ref{eq:lb-Dw}) will follow
from Lemma \ref{lemma:mut-info} if
\begin{equation}\label{eq:mut-inf-bd}
D_W(\varphi^{(n)}) \geq \sigma^2 (1-\rho^2) 2^{-\frac{2}{n} I({\bf S}_1;{\bf Y}|{\bf S}_2)}.
\end{equation}
To this end we denote by $R_W(D)$ the rate-distortion function for a
source of the law of ${\bf W}$. We then have 
\begin{align}
n R_W(D_W(\varphi^{(n)})) &\stackrel{a)}{\leq}
I({\bf W};\varphi^{(n)}({\bf S}_2,{\bf Y})) \nonumber \\ 
&\stackrel{b)}{\leq} I({\bf W};{\bf Y},{\bf S}_2) \nonumber \\
&= I({\bf S}_1-\rho {\bf S}_2;{\bf Y},{\bf S}_2) \nonumber \\
&= h({\bf S}_1-\rho {\bf S}_2) - h({\bf S}_1-\rho {\bf S}_2 | {\bf Y},{\bf S}_2) \nonumber \\
&\stackrel{c)}{=} h({\bf S}_1-\rho {\bf S}_2|{\bf S}_2) - h({\bf S}_1-\rho {\bf S}_2 |
{\bf Y},{\bf S}_2)\nonumber  \\
&= h({\bf S}_1|{\bf S}_2) - h({\bf S}_1|{\bf Y},{\bf S}_2) \nonumber \\
&= I({\bf S}_1;{\bf Y}|{\bf S}_2), \label{eq:mut-inf-Dw}
\end{align}
where inequality a) follows by the data-processing inequality and the
convexity of $R_W(\cdot)$. Inequality b) follows by the
data-processing inequality, and c) follows since ${\bf S}_2$ and ${\bf
  S}_1-\rho {\bf S}_2$ are independent.

Replacing $R_W(D_W(\varphi^{(n)}))$ in (\ref{eq:mut-inf-Dw}) by its
explicit form gives
\begin{displaymath}
\frac{n}{2} \log_2 \left(
  \frac{\sigma^2(1-\rho^2)}{D_W(\varphi^{(n)})} \right) \leq I({\bf
  S}_1;{\bf Y}|{\bf S}_2).
\end{displaymath}
Rewriting this inequality gives (\ref{eq:mut-inf-bd}), which combines
with Lemma \ref{lemma:mut-info} to  prove (\ref{eq:lb-Dw}).

\subsubsection{``Upper Bound'' on minimal $D_W(\varphi^{(n)})$}
We show that Assumption \ref{as:false} implies that the estimator 
\begin{align*}
\tilde{\varphi}^{(n)}({\bf S}_2,{\bf Y}) &= \alpha \cdot \hat{\bf S}_1
- \beta \cdot {\bf S}_2\\
&= \alpha \E{{\bf S}_1|{\bf Y}} - \beta {\bf S}_2,
\end{align*}
\begin{equation}\label{eq:alpha-beta}
\alpha \triangleq (1-\rho ) \frac{\sigma^2}{D^{\ast}} \qquad \text{and} \qquad
\beta \triangleq (1-\rho ) \frac{\sigma^2 -D^{\ast}}{D^{\ast}},
\end{equation}
violates (\ref{eq:lb-Dw}). To prove this we use the following two lemmas:
\begin{lm}\label{lemma:expectation-bound}
For any scheme achieving $D^{\ast}$ and any $\delta > 0$ there exists
an $n_0(\delta)$ such that for all $n \geq n_0(\delta)$ the following
three inequalities are satisfied
\begin{align}
\frac{1}{n} \sum_{i=1}^n \E{S_{1,k} \hat{S}_{1,k}} &\geq \sigma^2 -
D^{\ast} - \delta, \label{eq:corr-bound-S1-S1h} \\
\frac{1}{n} \sum_{i=1}^n \E{\hat{S}_{1,k}^2} &\leq \sigma^2-D^{\ast} +
\delta, \label{eq:corr-bound-S1h2}\\
\frac{1}{n} \sum_{i=1}^n\E{\hat{S}_{1,k} S_{2,k}} &\leq
\sigma^2-D^{\ast} + 2\delta. \label{eq:corr-bound-S2-S1h}
\end{align}
\end{lm}
\vspace{3mm}
\begin{lm}\label{lemma:non-neg-coeff}
For all $P/N \leq \rho/(1-\rho^2)$ we have
\begin{equation}\label{eq:non-neg-coeff}
\alpha (\rho - \beta) \geq 0.
\end{equation}
\end{lm}
\vspace{3mm}
The proofs of Lemma \ref{lemma:expectation-bound} and Lemma
\ref{lemma:non-neg-coeff} will be discussed in Section
\ref{subsec:proofs-lemmas}.  We now derive the desired upper bound on
$D_W(\tilde{\varphi}^{(n)})$
\begin{align}
D_W(\tilde{\varphi}^{(n)}) &=\frac{1}{n} \E{ \| {\bf W} -
  \tilde{\varphi}({\bf S}_2,{\bf Y}) \|^2} \nonumber\\
&= \frac{1}{n}
\sum_{k=1}^n \E{(S_{1,k}-\rho S_{2,k} -\alpha \hat{S}_{1,k} + \beta
  S_{2,k})^2} \nonumber\\
&= \frac{1}{n} \sum_{k=1}^n \E{(S_{1,k} -\alpha \hat{S}_{1,k} - (\rho
  - \beta ) S_{2,k})^2} \nonumber\\
&= \frac{1}{n} \sum_{k=1}^n \Big( \E{S_{1,k}^2} - 2 \alpha \E{S_{1,k}
  \hat{S}_{1,k}} \nonumber\\
&\qquad \qquad - 2(\rho-\beta ) \E{S_{1,k} S_{2,k}} + \alpha^2
\E{\hat{S}_{1,k}^2} \nonumber\\ 
&\qquad \qquad + 2\alpha (\rho - \beta) \E{\hat{S}_{1,k} S_{2,k}} \nonumber\\
&\qquad \qquad + (\rho-\beta)^2 \E{S_{2,k}^2}  \Big) \nonumber\\
&\leq \sigma^2 - 2 \alpha (\sigma^2 - D^{\ast} - \delta) -
2(\rho-\beta ) \rho \sigma^2 \nonumber\\
&\; \; \; \; + \alpha^2 (\sigma^2 - D^{\ast} + \delta) + 2\alpha (\rho
- \beta) (\sigma^2 - D^{\ast} + 2\delta) \nonumber\\
&\; \; \; \; + (\rho-\beta)^2 \sigma^2 \label{eq:achv-Dw-part1}
\end{align}
where the last step follows from Lemma \ref{lemma:expectation-bound},
using the fact that $\alpha \geq 0$, and using Lemma
\ref{lemma:non-neg-coeff}.

Upon letting $n$ tend to infinity, we obtain
\begin{align*}
\varlimsup_{n \rightarrow \infty} D_W(\tilde{\varphi}^{(n)}) &\leq
\sigma^2 - 2 \alpha (\sigma^2 - D^{\ast} - \delta) -
2(\rho-\beta ) \rho \sigma^2 \\
&\; \; \; \; + \alpha^2 (\sigma^2 - D^{\ast} + \delta)\\
&\; \; \; \; + 2\alpha (\rho - \beta) (\sigma^2 - D^{\ast} + 2\delta)
+ (\rho-\beta)^2 \sigma^2.
\end{align*}
But since $\delta >0$ was arbitrary,
\begin{align*}
\varlimsup_{n \rightarrow \infty} D_W(\tilde{\varphi}^{(n)})
&\leq \sigma^2 - 2 \alpha (\sigma^2 - D^{\ast}) -
2(\rho-\beta ) \rho \sigma^2\\
&\; \; \; \; + \alpha^2 (\sigma^2 - D^{\ast}) + 2\alpha (\rho
- \beta) (\sigma^2 - D^{\ast} )\\
&\; \; \; \; + (\rho-\beta)^2 \sigma^2\\
&\stackrel{a)}{\leq} \sigma^2 (1-\rho) \left( 2 - \frac{\sigma^2}{D^{\ast}}
  (1-\rho) \right)\\
&\stackrel{b)}{<} \sigma^2 (1-\rho) \left( 2 -
  \frac{N+2P(1+\rho)}{N+P(1-\rho^2)} (1-\rho) \right)\\
&= \sigma^2 (1-\rho^2) \frac{N}{N + 2P(1-\rho^2)},
\end{align*}
which contradicts (\ref{eq:lb-Dw}). Here, a) follows from
(\ref{eq:alpha-beta}), and b) since we assumed $D^{\ast} < D_u$.
\hspace{5cm}$\Box$

\subsection{Proofs of Lemmas}\label{subsec:proofs-lemmas}

To prove Lemma \ref{lemma:mut-info} we first notice that the
assumption $P/N \leq \rho/(1-\rho^2)$ implies, by (\ref{eq:Rd-S1-S2})
\& (\ref{eq:uncoded}), that
\begin{displaymath}
R_{S_1,S_2}(D_u,D_u) = \frac{1}{2} \log_2 \left( 1 +
  \frac{2P(1+\rho)}{N} \right).
\end{displaymath}
Hence,
\begin{align}
\frac{n}{2} \log_2 \left( 1 + \frac{2P(1+\rho)}{N} \right)
&= n R_{S_1,S_2}(D_u,D_u) \nonumber \\
&\stackrel{a)}{<} n R_{S_1,S_2}(D^{\ast},D^{\ast}) \nonumber \\
&\stackrel{b)}{\leq} \sum_{k=1}^n I(X_{1,k},X_{2,k} ;
Y_k) \qquad \qquad \nonumber \\
&\stackrel{c)}{\leq} \frac{n}{2} \log_2 \left( 1 +
  \frac{2P(1+\tilde{\rho})}{N}\right)\label{eq:RD-sum-rate}
\end{align}
where $\tilde{\rho}$ is given in (\ref{eq:tilde-rho}). Here a) follows
from the assumption $D^{\ast} < D_u$ and the strict monotonicity of
$R_{S_1,S_2}(D,D)$; b) follows from (\ref{eq:prf-sum-rate1})
\& (\ref{eq:prf-sum-rate2}); and c) follows from Lemma
\ref{lemma:MAC-rate}. From (\ref{eq:RD-sum-rate}) and
(\ref{eq:tilde-rho}) we conclude that
\begin{equation}\label{eq:min-corr-X1X2}
\frac{\left| \frac{1}{n} \sum_{k=1}^n
    \E{X_{1,k}X_{2,k}} \right|}{\sqrt{\left( \frac{1}{n} \sum_{k=1}^n
      \E{X_{1,k}^2} \right) \left( \frac{1}{n} \sum_{k=1}^n
      \E{X_{2,k}^2} \right)}} > \rho.
\end{equation}
The lemma now follows from \eqref{eq:prf-side-info1-2},
Lemma \ref{lemma:MAC-rate} inequality \eqref{eq:MAC-rate1}, and
(\ref{eq:min-corr-X1X2}). \hspace{7.1cm}$\Box$


We turn to Lemma \ref{lemma:expectation-bound} and begin by proving
Inequalities (\ref{eq:corr-bound-S1-S1h}) and
(\ref{eq:corr-bound-S1h2}). By the definition of achievability, for
any scheme achieving $D^{\ast}$ and any $\delta > 0$ there must exist
an $n_0(\delta)$ such that for all $n \geq n_0(\delta)$ 
\begin{equation}\label{eq:distortion}
D^{\ast} - \delta < \frac{1}{n} \sum_{k=1}^n \E{(S_{i,k} -
  \hat{S}_{i,k})^2} < D^{\ast} + \delta \qquad i=1,2.
\end{equation}
Since, by our assumption that $\hat{\bf S}_1 = \E{{\bf S}_1|{\bf Y}}$, the
orthogonality principle must be satisfied,
we obtain from (\ref{eq:distortion}) that
\begin{align}
\sigma^2 - D^{\ast} -\delta &\leq \frac{1}{n} \sum_{k=1}^n
\E{S_{1,k}\hat{S}_{1,k}} \leq \sigma^2 - D^{\ast} +\delta,
\label{eq:bound-corr-S1S1h} \\
\sigma^2 - D^{\ast} -\delta &\leq \frac{1}{n} \sum_{k=1}^n
\E{\hat{S}_{1,k}^2} \leq \sigma^2 - D^{\ast} + \delta. \label{eq:bound-corr-S1h2}
\end{align}

To prove the inequality (\ref{eq:corr-bound-S2-S1h}) we start by
observing that any scheme achieving $D^{\ast}$ must satisfy
\begin{equation}\label{eq:symmetry}
D_1 = D_2 = D^{\ast}.
\end{equation}
This follows by a time-sharing argument: assume there would
exist a scheme achieving $D^{\ast}$ with $D_1 = D^{\ast}$
and $D_2 = \tilde{D} < D^{\ast}$. Then, by symmetry there would also
exist a scheme achieving $D^{\ast}$ with $D_1 = \tilde{D} < D^{\ast}$
and $D_2 = D^{\ast}$. Time-sharing between those two schemes
would give a scheme achieving $1/2(D^{\ast}+\tilde{D}) <
D^{\ast}$ which contradicts the definition of $D^{\ast}$.

Statement (\ref{eq:symmetry}) implies, in view of
(\ref{eq:distortion}), that for any scheme achieving $D^{\ast}$ and
any $\delta > 0$ there must exist an $n_0(\delta)$ such that for all
$n \geq n_0(\delta)$
\begin{displaymath}
\frac{1}{n} \sum_{k=1}^n \E{(S_{2,k}-\hat{S}_{1,k})^2} \geq
\frac{1}{n} \sum_{i=1}^n \E{(S_{1,k}-\hat{S}_{1,k})^2} -2\delta,
\end{displaymath}
which is equivalent to
\begin{equation}\label{eq:corr-S2-S1h}
\frac{1}{n} \sum_{k=1}^n \E{S_{2,k} \hat{S}_{1,k}} \leq \frac{1}{n}
\sum_{k=1}^n \E{S_{1,k} \hat{S}_{1,k}} + \delta.
\end{equation}

Applying (\ref{eq:bound-corr-S1S1h}) to the RHS of
(\ref{eq:corr-S2-S1h}) gives
\begin{displaymath} 
\frac{1}{n} \sum_{k=1}^n \E{S_{2,k}\hat{S}_{1,k}} \leq \sigma^2 -
D^{\ast} +2\delta.
\end{displaymath}
\hspace{8.5cm}$\Box$

To prove Lemma \ref{lemma:non-neg-coeff} we notice that $\alpha$ is
always positive. Hence, the proof of Lemma \ref{lemma:non-neg-coeff}
merely requires showing $\beta \leq \rho$ whenever $P/N \leq
\rho/(1-\rho^2)$. Furthermore, since $D^{\ast}$ is certainly 
non-increasing in $P/N$, and therefore $\beta$ is non-decreasing in
$P/N$, it is sufficient to show that $\beta \leq \rho$ for $P/N =
\rho/(1-\rho^2)$. And this follows from plugging the lower bound
(\ref{eq:one-point}) for $D^{\ast}$ in the expression for $\beta$.
\hspace{2.2cm}$\Box$

\end{document}